# Griffiths like Robust Ferromagnetism in $Co_{3-x}Mn_xTeO_6$; ($x$ = 0.5, 1, 2)

**Harishchandra Singh**[1,2,♣], Haranath Ghosh[1,2], C. L. Prajapat[3], and M. R. Singh[3]

[1]Homi Bhabha National Institute

[2]Indus Synchrotron Utilization Division

Raja Ramanna Centre for Advanced Technology, Indore-452013, India

[3]Technical physics division, Bhabha Atomic Research Centre, Trombay, Mumbai-400085, India

Dated: October 17, 2015

We report near room temperature ferromagnetic (FM) as well as low temperature antiferromagnetic (AFM) correlations in $Co_{3-x}Mn_xTeO_6$ (CMTO); ($x$ = 0.5, 1 and 2) solid solutions using thorough DC magnetization studies. For all $x$, CMTO show not only short-range robust FM order ~ 185 K, but also long range enhanced AFM order ≤ 45 K. Griffiths-like FM interactions show up in magnetization data; it exists over an extended temperature range and remains unsuppressed in magnetic field as large as 1 T, indicating its robustness. Variations in both FM and AFM phases as a function of Mn concentration also support our observation of anomalous behavior in average bond distances and charge states (JAP 116: 074904 (2014)). Further, an attempt towards the structural insight into the observed complex magnetic behavior by using network like structural analysis has been drawn. These observations make CMTO an interesting magnetic system from fundamental and application perspective.

**Keywords:** A. Ceramics; B. Chemical synthesis; C. X-ray diffraction; D. Magnetic properties

## I. INTRODUCTION

Coexistence of magnetic interactions of opposite nature leading to complex magnetic behavior has been seen in numerous magnetic systems, for example in manganites and cobaltites owing to their enormous potential applications in different fields [1-4]. Griffiths phase (GP), first proposed by Griffith [5-8], presents the competition between the ferromagnetic (FM) and antiferromagnetic (AFM) interactions under random potential [5]. Disorder sets in such a way that different values of exchange coupling may be assigned randomly to different sites of the lattice. This causes existence of short-range FM clusters when $T_C < T < T_{GP}$ (here, $T_C$ represents

[♣] Corresponding author. Tel.: +91-731-244 2583; Fax: +91-731-244 2140.
E-mail address: singh85harish@gmail.com, singh85harish@rrcat.gov.in

the FM Curie temperature, whereas, $T_{GP}$ is the Griffiths temperature at which the FM clusters begin to nucleate). The intermediate regime is known as GP. The Hallmark of GP is a sharp downturn in the inverse magnetic susceptibility (as a function of temperature) [5-8]. This downturn in the thermal behavior of $\chi^{-1}$ is an important observation that distinguishes GP from smeared phase transition because, the latter gives rise to an upward curvature in $\chi^{-1}$ vs T above $T_C$, deviating from Curie - Weiss (CW) law [5-8]. The softening of the downturn in $\chi^{-1}$ with the progressive increase in field is another typical signature of GP. The basic characteristics of GP regime is that above $T_C$ there exists finite but nano size clusters with FM correlated spins [3-5].

Role of Griffiths like ferromagnetism (well accepted phenomenon in CMR: Colossal Magneto Resistance [1-4]) in multiferroic (MF) systems cannot be overemphasized. We show that presence of such a phase in multiferroics can add to a great advantage. MFs (exhibit either coupling between electronic and magnetic orders (type II) or a separate single order (type I)), offer great opportunities for applications in information storage, spin electronics, magneto-electronics and solar cells. A number of MFs possessing cationic and anionic non-stoichiometry are further essential in many other applications such as energy conversion, oxygen sensing, oxygen storage etc.[9]. During last few decades, type II MF materials which are of practical interest are found only at very low temperatures except a few [10]. Besides understanding on the mechanism of coupling of these ferroic orders, designing and finding new MF materials are some of the frontier research activities [11,12]. Undoubtedly, controlling and coupling of various 'ferro or antiferro' magnetic orders at room temperature are of immense interest [10-14]. Enhancement of these couplings at much higher temperatures may possibly be achieved by various ways like internal chemical pressure (through doping) and other external perturbations etc. [15,16]. Many of the doped compounds become intrinsically inhomogeneous due to random distribution of cation sizes, their different valence/spin states and strong competition between different ordering tendencies and present great advantage in the field [15,16]. Compounds on doping, not only show the enhanced magnetoelectric coupling at higher temperature but also coexistence of more than one magnetic interaction, e. g. coexistence of AFM and FM. [15-17].
$Co_3TeO_6$ (CTO), a low symmetry (monoclinic: C*2/c*) type II MF at low temperatures, shows complex magnetic structure with a sequence of AFM transitions [18]. $Mn_3TeO_6$ (MTO), on the other hand, crystallizes in higher symmetry (Rhombohedral: R$\bar{3}$) type I multiferroic, exhibits similar AFM transition at low temperature [19]. Structural details can be found elsewhere [18,19]. MTO and CTO show AFM transitions ($T_N$, the Neel temperature) at around 23 K and 26 K, respectively [18,19]. In contrast, Mn doped CTO {($Co_{3-x}Mn_xTeO_6$ (CMTO); ($x$ = 0.5, 1 and 2)} or vice-versa Co doped MTO not only enhances the AFM transition temperature [20,21] but also show possible high temperature FM correlations (for $x$ = 0.5) [17]. Our recent structural and spectroscopic studies suggest possible structural origin of this enhancement in AFM transition of CMTO [17], which was not understood earlier [20,21]. For $x \geq 0.5$, increase in lattice parameters and average transition metal - oxygen (TM – O) bond distances with increasing Mn concentration is corroborated with the magnetic behavior of enhanced AFM transition temperatures and is established. Together with these, relative ratios of $Co^{3+}/Co^{2+}$ and $Mn^{3+}/Mn^{2+}$ (or $TM^{3+}/TM^{2+}$), which has been observed through XANES analysis at Co and Mn K-edges, found to be maximum at around $x \sim 0.5$. At this particular concentration, average <Co/Mn-O> bond lengths are found to be lowest. Approximately at the same concentration, maximum $T_N \sim$ 45 K has been observed and could be correlated with the structural and spectroscopic results, which show anomalous behavior in average bond lengths and charge ratios at $x \sim 0.5$ [17]. In this report, on the other hand, we describe the probable mechanism behind the observance of high temperature FM correlations (along with enhanced AFM ordering) in these solid solutions, not reported so far.

## II. EXPERIMENTAL DETAILS

Single phasic polycrystalline CMTO ($x$ = 0.5, 1 and 2) solid solutions were synthesized using conventional solid state reaction route [17]. The reactants were used as commercial cobalt oxide $Co_3O_4$ (Alfa Easer 99.7 %), $Mn_3O_4$ (Alfa Easer 99.99%) and tellurium dioxide $TeO_2$ (Alfa Easer 99.99 %). The ground oxide mixtures were calcined at 700 $^o$C for 10 hrs and then recalcined at 800 $^o$C for ~ 25 hrs as a second step. For each step of calcinations and sintering (~ 850 $^o$C for 2hrs), the pellets were made by applying 2 tons of pressure. Thorough structural and spectroscopic characterizations were performed using Synchrotron X-ray at Indus-2 [17]. DC magnetization measurements were carried out using a SQUID magnetometer (MPMS5 by Quantum Design).

## III. RESULTS AND DISCUSSION

### A. Magnetic behavior of $Co_{3-x}Mn_xTeO_6$ ($x$ = 0.5, 1 and 2)

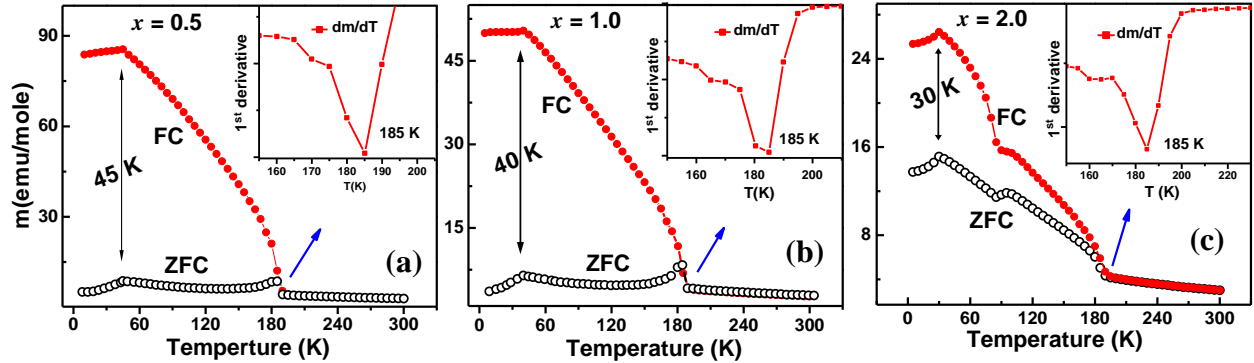

**Fig. 1.** (a), (b), (c) Low magnetic field DC magnetization in ZFC/FC protocol (taken at 100 Oe) for $x = 0.5$, 1.0 and 2.0 indicates AFM transitions at 45 K, 40 K and 30 K, respectively followed by ferromagnetic like transition at around 185 K. Insets show the dm/dT to assign exact high temperature transition.

Fig. 1 (a) [17], (b) and (c) show temperature dependent magnetization (M vs T) for CMTO ($x = 0.5$, 1.0, and 2.0) solid solutions performed at a constant magnetic field of 100 Oe. All these DC magnetization data have been recorded under both the zero field cooled (ZFC) and field cooled (FC) conditions, as shown in Fig. 1. We have also used constant magnetic fields of 10 kOe and 50 kOe during the magnetization measurements which will be discussed further. It is interesting to note that the observed magnetic behavior for all the composition is similar (Fig. 1) but much richer than those reported earlier [20-21]. Looking from room temperature side, in Fig. 1, samples exhibit interestingly paramagnetic (PM) to FM like transition at a characteristic temperature $T_C$ (Curie temperature) ~ 185 K, which is same (within ± 2 K) for all the samples, and is the main focus of the present report. This feature in the magnetization curves is followed by the AFM transitions at $T_N$ ~ 45 K, 40 K and 30 K (as shown by double headed arrow in both ZFC and FC curves), for 0.5, 1.0 and 2.0 Mn content, respectively. However, in literature [20-21], the only reported transitions for these solid solutions were AFM like and at temperatures ≤ 40 K. These values (reported earlier ~ 40 K and observed presently ~ 45 K) are much higher compared to the Neel temperatures observed in either of the compounds CTO (26 K) and MTO (23 K) separately [18,20,21]. We have corroborated this enhancement with our earlier structural and spectroscopic results, which is explained via variations in average bond lengths and average charge ratios as a function of Mn concentration [17].

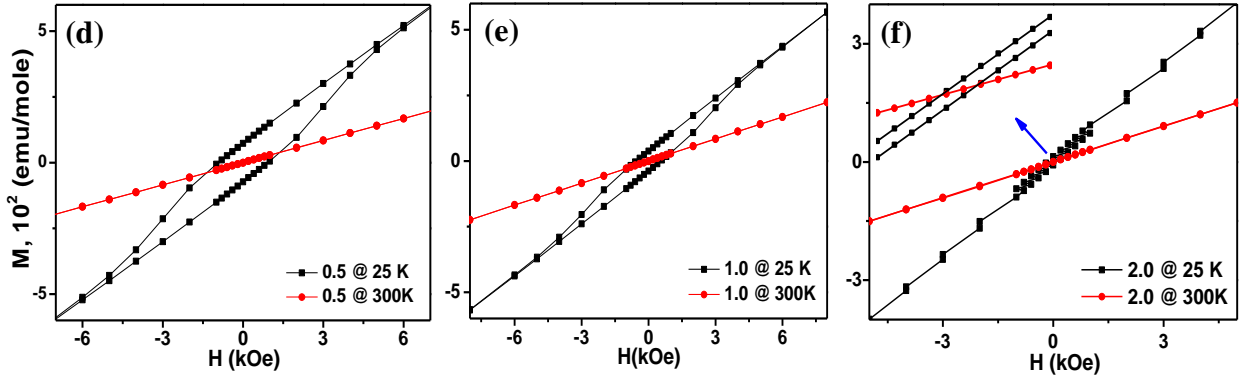

**Fig. 2.** (d), (e), (f) Corresponding magnetic hysteresis (M/H) measurements for *x* = 0.5, 1.0 and 2.0 at 300 K and 25 K signify the paramagnetic behavior at room temperature and FM like at low temperature.

Further, to the best of our knowledge, there is no FM correlation and hence corresponding transition reported in these compounds [20-21]. We, on the other hand, observe a FM like transition at T ~ 185 K along with a bifurcation in ZFC and FC curves for all the samples (see Figs. 1 (a), (b) and (c)). In literature [22-23], the bifurcation in FC–ZFC magnetization data has been attributed to the presence of a spin-glass, cluster-glass, super-PM behaviour etc. We attribute this feature to a PM to FM like transition. As discussed earlier in several other reports,[22,23] the temperature derivative dm/dT of the FC magnetic data visualizes the FM transition, and is shown in the inset of Fig. 1 (a), (b) and (c). The existence of such type of FM correlations (below $T_N$) in the AFM ground state, reported earlier for various other compounds [6-8], have been understood as the presence of FM interaction in the AFM phase, and is demonstrated below. The statement made during the analysis of magnetization data, is confirmed through the magnetic hysteresis data (M vs H), as can be seen in Fig. 2 (d), (e) and (f). The PM state is confirmed by the reversible linear M-H curve at 300 K, and the persistence of FM nature from below 185 K down to 10 K is revealed by the presence of M-H loops. The loop does not show any signature of saturation up to 10 kOe which indicate presence of significant AFM contribution. In addition, the magnetic hysteresis loop at 25 K signifies the existence of ferromagnetism even below $T_N$ suggesting the coexistence of FM and AFM phases.

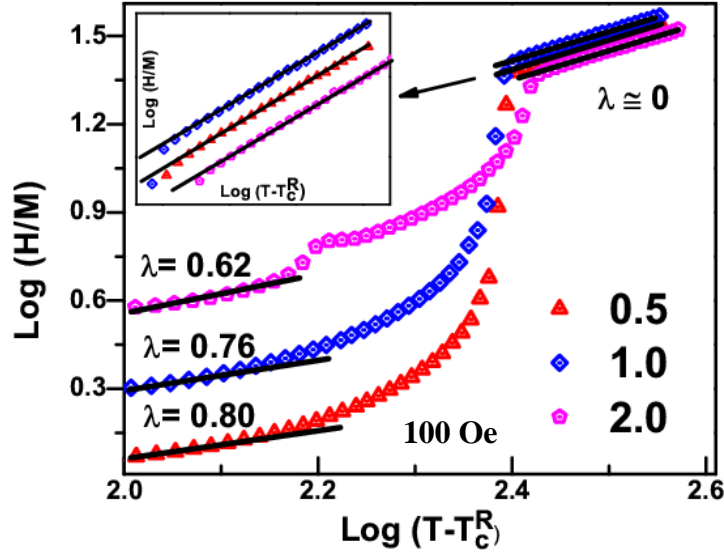

**Fig. 3.** FM correlations is manifested through the CW fit of the inverse magnetic susceptibility (H/M) data for CMTO (*x* = 0.5, 1 and 2) solid solutions, inset enlarge the linear part. Data has been offset along Y-axis for clarity.

**Table I.** Effective Bohr magneton ($\mu_{eff}$) and $\lambda$ values for both GP and PM phase obtained from fitted parameters (observed through CW fit) and modified CW law for CMTO. Maximum errors in $\mu_{eff}$ and $\lambda$ are 0.06 and 0.005, respectively.

| *x* = | 0.5 | | | 1.0 | | | 2.0 | | |
|---|---|---|---|---|---|---|---|---|---|
| H (Oe) | 100 | 10000 | 50000 | 100 | 10000 | 50000 | 100 | 10000 | 50000 |
| $\mu_{eff}$ ($\mu_B$/f.u.) | 8.97 | 9.14 | 9.25 | 9.03 | 9.22 | 9.33 | 9.42 | 9.67 | 9.83 |
| $\lambda_{PM}$ | 0.002 | 0.003 | 0 | 0.004 | 0.003 | 0 | 0.003 | 0.004 | 0 |
| $\lambda_{GP}$ | 0.80 | 0.29 | - | 0.76 | 0.244 | - | 0.62 | 0.145 | - |

Presence of FM correlations is also manifested through the CW fit of the inverse magnetic susceptibility $\chi^{-1}$ (H/M) data (see Fig. 3). To investigate the same, we analyses M vs T data in detail. The normal trend to analyze the M vs T data is through CW fit, which may provide the nature of existing magnetic interactions and effective magnetic moment [18]. Fit shows negative value of $\Theta_{CW}$ (CW parameter) suggesting presence of strong AFM interactions in CMTO, in agreement with earlier reports [20-21]. The effective magnetic moment ($\mu_{eff}$) is estimated from the CW parameter (obtained from $\chi^{-1}$ = (T- $\Theta_{CW}$)/$C_{CW}$) using $\mu_{eff}/f.u. = \sqrt{(8 C_{CW})}$ relation. It can be seen that all the $\chi^{-1}$ vs T curves follow CW law just above $T_{GP}$ ~190 K (onset of downturn temperature), with an effective magnetic moment, $\mu_{eff}$ ~ 8.97 - 9.42 $\mu_B$ / *f. u.* and PM negative Curie temperature of $\Theta_{CW}$ ~ 49.13 K to 68.85 K for $x$ = 0.5 to 2.0, respectively. Both values ($\mu_{eff}$ and $\Theta_{CW}$) are significantly larger than the previous reports of $\Theta_{CW}$ (-ve) which range from about 35 K to 45 K and $\mu_{eff}$ ~ 5.8 - 6.0 $\mu_B$ / $f. u.$ [20-21]. In table I, we tabulate all the fitted parameters observed from CW fit for all $x$ values. As we go from $x$ = 0.5 to $x$ = 2.0, $\Theta_{CW}$ increases while $T_N$ decreases, indicating weakening of the AFM interactions on Mn doping. Interestingly, the observed larger magnetic moment, which increases with Mn concentration, indicates FM cluster type interactions in our samples [24]; this is demonstrated further.

Evidence of short range FM correlation is also presented through the down turn in the (log-log plot) inverse magnetic susceptibility data (a hallmark of GP) [6-8] and magnetization difference curve [25], and are shown in Figs. 3 and 4, respectively. Fig. 3 represents one of the main characteristics of GP i.e., it follows power law: $\chi^{-1} = (T-T_C^R)^{1-\lambda}$, where $0 < \lambda < 1$ and $T_C^R = T_N$, respectively. The $\lambda$ values for paramagnetic ($\lambda_{PM}$) and Griffiths phase ($\lambda_{GP}$) are tabulated in table I. We find $\lambda_{GP}$ increases with increasing $x$ from 0.62 to 0.80 for $x$ = 0.5 to $x$ = 2.0. These values of $\lambda_{GP}$ correspond to strong FM correlation in CMTO [6-8]. One of the possible reasons for this ferromagnetic ordering is $TM^{2+}$- O - $TM^{3+}$ networks (as a result of the local appearance of $TM^{3+}$ in $TM^{2+}$-O-$TM^{2+}$ networks). This is may be because of coexistence of mixed oxidation states (+2 and +3) of Co and Mn, as described in our previous study [17], and the fact that unlike spins (i.e. network corresponding to different charge states) favor AFM interactions while like spins favor FM interactions [26-27].

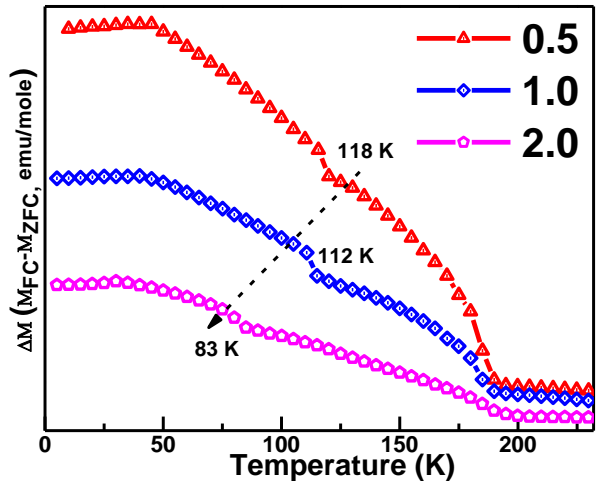

**Fig. 4.** Magnetization difference curve as a function of temperature for CMTO ($x$ = 0.5, 1 and 2). This is just to show additional transitions at 118 K, 112 K and 83 K.

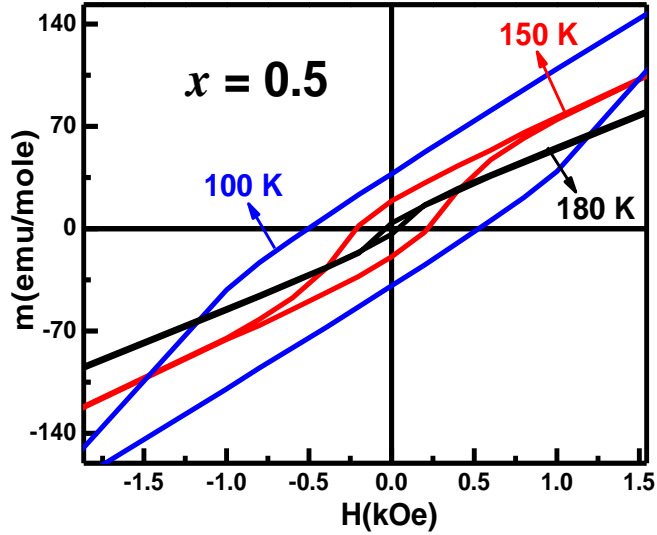

**Fig. 5.** Representative hysteresis loop for CMTO; $x = 0.5$ at 180 K, 150 K and 100 K.

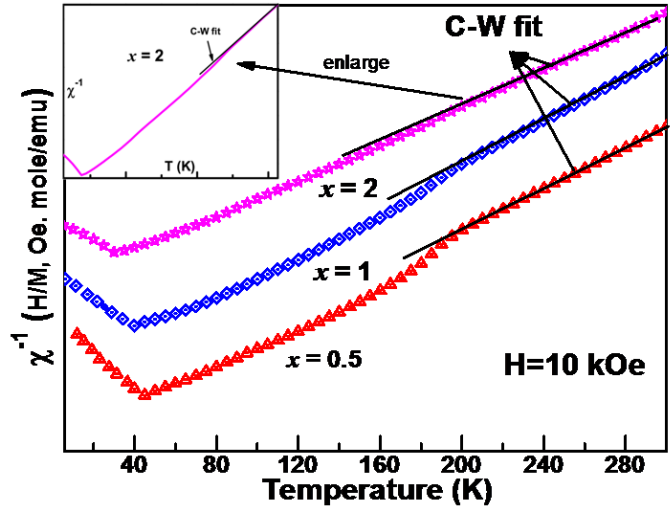

**Fig. 6.** Deviation of H/M curve from the CW law, the upper inset shows the enlarged data for $x = 2$. Main panel data is shifted in Y axis for clarity, corresponding values are not

Furthermore, the FM correlation is also visible from the magnetization difference ($M_{FC}$-$M_{ZFC}$) versus temperature plot. Fig. 4 shows the temperature-dependent $\Delta M = M_{FC}$-$M_{ZFC}$ curves at 100 Oe. This subtraction advantageously eliminates PM and diamagnetic contributions and simultaneously indicates the presence of hysteresis (if $\Delta M \neq 0$) [25]. The FM characteristic transition temperature $T_{FM}$ may be obtained by taking $\frac{d}{dT}(M_{FC} - M_{ZFC})$ from the ($M_{FC}$-$M_{ZFC}$) versus temperature curve. Fig. 4 shows few additional anomalies at around 118 K, 112 K and 85 K for $x = 0.5, 1.0$ and $2.0$, respectively. These transitions were hidden in M vs T data, which can be seen clearly through ΔM vs T curve. Representative magnetic hysteresis loops at 180 K, 150 K and 100 K for $x = 0.5$ (Fig. 4) indicate FM nature of these transitions. This additional hidden magnetic transitions may be due to the several combinations of possible $TM^{2+}$-O-$TM^{2+}$, $TM^{3+}$-O-$TM^{2+}$, $TM^{3+}$-O-$TM^{3+}$ networks [26-27]. Moreover, the strength of FM correlation shows suppression with increase in magnetic field, similar to GP like [6-8].

To further examine whether the short-range FM correlations can be ascribed to GP; we have measured M vs T data and have plotted $\chi^{-1}$ (H/M) vs T for 100 Oe, 10 kOe and 50 kOe, as shown in Figs. 6 and 7. According to GP, the downturn feature in $\chi^{-1}$ is expected to increase with decreasing field strength, at least for low H, where the susceptibility of the clusters is dominant. This behavior is clearly reflected in Fig. 6 (and Fig. 3). At higher fields, the contribution from PM matrix is significant, as a result, $\chi^{-1}$ vs T curve becomes almost linear in GP region, as shown in Fig. 7. Similar downward turn (and the corresponding suppression) in $\chi^{-1}$ (T) has also been observed in several other compounds, and attributed to GP, owing to the formation of nano size FM domains [6-8]. Formation of nano-size Griffiths-like FM clusters has also been demonstrated by small-angle neutron scattering [28]. Often, the same is reflected in the very large value of effective Bohr magneton, $\mu_{eff}$ as discussed above.

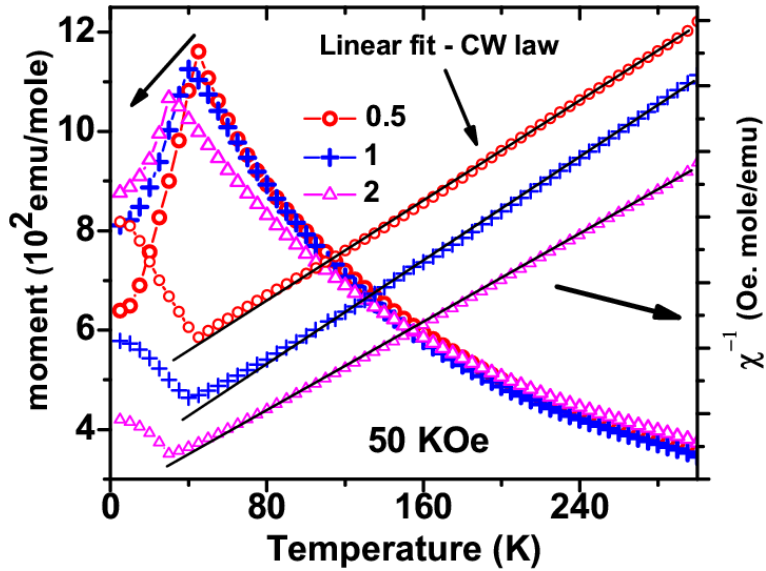

**Fig. 7.** Very high magnetic field χ (H/M) and $\chi^{-1}$ curve for CMTO, which signifies another characteristic of GP. The CW fit indicates absence of GP at 50 kOe.

So far we have illustrated the signatures of FM correlations along with its short-range nature in CMTO ($x = 0.5$, 1 and 2) solid solutions using M vs T and M vs H studies. This feature has been attributed to the FM phase embedded in AFM matrix, which is of Griffith like. This is quite unusual because the same was not observed in previous investigations of CMTO [21-21]. We attribute this GP like anomaly to the presence of mixed valence TM ions. Also, as the Mn content is changed, it modifies the H/M vs T curves very differently and changes the value of $\lambda_G$ (Table I). The downturn of the H/M curves show suppression, whereas there is no change in AFM transition temperatures ($T_N$), as shown in Fig. 8, with the increase in the Co/Mn ratio. Our observations also support the absence of ferromagnetism in MTO, whereas a very weak presence of FM interactions in CTO [78,96]. In the following, we study (the magnetic networks like) structural study in order to understand interesting magnetic behaviour of the CMTO samples.

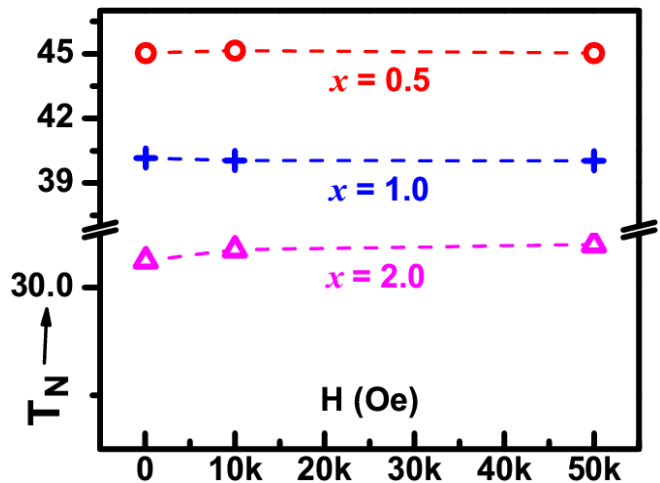

**Fig. 8.** Unaltered values of $T_N$ at various magnetic fields for CMTO ($x = 0.5$, 1 and 2).

**B. Structural insight for the observed magnetic behavior**

We present detailed structural aspects in order to find its possible correlation to magnetic behavior presented in the preceding paragraphs; in connection with earlier neutron diffraction study [20-21]. Fig. 9 shows representative layer and equilateral triangular arrangements of Mn/Co (ions structure) of studied CMTO solid solution at room temperature for $x = 0.5$. Oxygen is not shown for the sake of clarity. First we show the polyhedral view of a representative R-3 structure for $x = 0.5$ (Fig. 9 (a)). Simplified unit cell of CMTO solid solution is displayed through Mn/Co ions in Fig. 9 (b). From the side view along *a*- or *b*- axis, a layered structure is observed (Fig. 9 (b)), and from the top view along *c* axis, an equilateral triangular arrangement of Mn/Co ions can be found (Fig. 9 (b)). Fig. 9 (c) (slightly tilted from Fig 9 (b)) show that the equilateral triangles (length = 3.699 Å) arranged by Mn ions. Few triangles are shown.

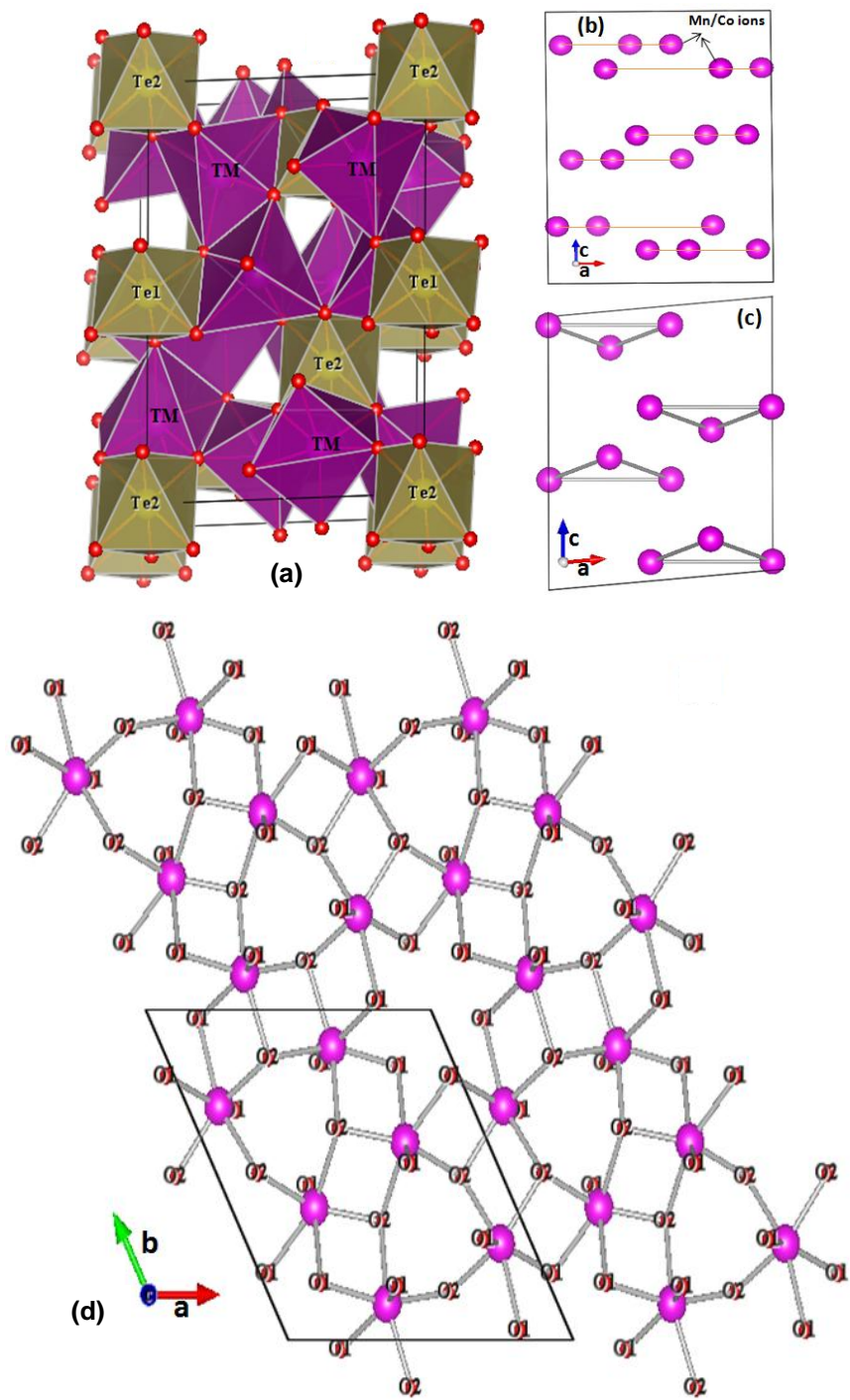

**Fig. 9.** (a) Polyhedral view along (010), (b) layered structure (side view from *b*-axis) of Mn/Co ions, (c) equilateral triangular arrangements (top view along *b* axis) of Mn/Co ions located at different layers, and (d) network type of structure through oxygen along *c*-axis (a-b plane).

These belong to different layers. Furthermore, a network structure is also presented (Fig. 9 (d)) along c-direction (a-b plane). Above analysis indicates that there are triangular cycles, which may be comprised of Mn/Co/Mn and Co/Mn/Co sites, stacked along $c$ axis of the unit cell, very similar to the earlier observation by Ivanov et al., [20-21] predicted using neutron diffraction study. The connectivity of TM or Mn/Co cations with increasing Mn concentration shows a deviation from the ideal triangular lattice, distortions occurred due to displacements of TM ion positions from their initial site [17,20,21]. The same has been seen in the previous investigation, during the Rietveld refinement of CMTO for $x \geq 0.5$ [17]. Further, to have a view of the above observations from neutron diffraction study; particular kind of triangular structure is shown in Fig 9 (d). It shows the possible cationic arrangements for this geometry along $c$ axis, which can extend throughout the lattice through the periodic translation. They are attached through O atoms which prefer to have either super exchange or double exchange interaction. Our earlier report [17] (along with mixed valence TM ions) clearly elaborates the increasing nature of average bond distances and distortion in the corresponding lattice decreases the magnetic interactions in these CMTO solid solutions.

We have discussed and demonstrated the origin of a Griffiths-like anomaly in CMTO solid solutions. Comparison of the present CMTO samples with that reported by Ivanao at. el, and Mathew et al., [20-21] give a possible route for double exchange phenomenon between various arrangements of $Co^{2+/3+}$ and $Mn^{2+/3+}$, wherein the combination of unlike spins often favor FM interactions. As discussed above, charge ratios of TM ions show maximum values at lower concentration and lower values for higher Mn concentration, which may explains the observance of such robust FM interactions in CMTO. Further, presence of $Co^{3+}/Mn^{3+}$ in the triangular lattice of Mn/Co/Mn and Co/Mn/Co may lead to (structural/magnetic) frustration and hence, spontaneously induce the FM and or AFM interaction, apart from the major AFM interaction proposed earlier [20-21]. This may cause disorder / magnetic inhomogeneity in CMTO samples leading to GP like phenomenon [6-8,22,31,32]. Triangular lattice usually sets in AFM arrangements with 120º alignment of spins, this along with inter-layer coupling and high/low spin states of $Co^{3+}/Mn^{3+}$ leads to competition between FM and AFM interactions. There are several other compounds which show similar magnetic behavior and have been explained using double exchange between mixed valence TM ions [33-35]. In the structural section, we speculate all the possible cause related to competing FM and AFM interactions. Moreover, JT distortion (due to $TM^{3+}$), on the other hand, can also induce GP like FM [6-8,22,31,32]. However, we do not consider the same in the present study because, out of the three known mode of octahedral distortions (breathing, basal plane and stretching modes), breathing mode (which cannot be responsible for GP) is found in our CMTO samples [17].

## IV. CONCLUSIONS

We identify CMTO ($x$ = 0.5, 1 and 2) solid solutions as a potential magnetic materials having very high temperature FM interactions, low temperature enhanced AFM interactions and their coexistence. This is demonstrated through detailed magnetization (M vs T and M vs H) studies. Scaling of inverse magnetic susceptibility data provide clear indication of Griffiths like FM phase extended over large thermal region. This FM phase is robust against magnetic field. Both the FM and AFM phases are sensitive to the content of Co/Mn ratio. We also attempt detailed structural aspects in order to find its possible correlation to the magnetic behavior possessed by CMTO solid solutions; in conjunction with earlier neutron diffraction study. We believe with our detail structural, spectroscopic and magnetic studies that such high temperature FM and low temperature enhanced AFM transition temperatures in CMTO would attract further research with an aim achieving room temperature FM and low temperature AFM together with possible dielectric coupling.

## ACKNOWLEDGMENTS

Mr. Harishchandra Singh thanks Dr. A. K. Sinha for his guidance and support during the work. HS also acknowledge M. N. Singh for his help during synchrotron XRD measurements. Authors acknowledge Dr. P. D. Gupta and Dr. P. A. Naik for their supports and encouragements. Mr. Singh also acknowledges Homi Bhabha National Institute, India for research fellowship.